\documentclass[twocolumn]{emulateapj}
\usepackage{graphicx}
\usepackage{epsfig}
\usepackage{dcolumn}
\usepackage{rotating}

\def\arcsec{\hbox{$^{\prime\prime}$}}

\begin{document}

\title{ The Incidence of Active Galactic Nuclei in Pure Disk Galaxies: 
The {\it Spitzer} View}

\author{S. Satyapal\altaffilmark{1}, T. B\"oker\altaffilmark{2}, W. Mcalpine\altaffilmark{1}, M. Gliozzi \altaffilmark{1}, N. P. Abel\altaffilmark{3}, \& T. Heckman\altaffilmark{4}}

\altaffiltext{1}{George Mason University, Department of Physics \& Astronomy, MS 3F3, 4400 University Drive, Fairfax, VA 22030; satyapal@physics.gmu.edu}

\altaffiltext{2}{ESA/ESTEC, Keplerlaan1, 2200AG Noordwijk, Netherlands}

\altaffiltext{3}{Department of Physics, University of Cincinnati, Cincinnati, OH 45221}

\altaffiltext{4}{Center for Astrophysical Sciences, Department of Physics and Astronomy, The Johns Hopkins University, Baltimore, MD 21218}

\begin{abstract}

Using the {\it Spitzer} telescope, we have conducted a high-resolution 
spectroscopic study of 18 bulgeless (Hubble type of Sd or Sdm) galaxies that 
show no definitive signatures of nuclear activity in their optical spectra. 
This is the first systematic mid-infrared search for weak or hidden active 
galactic nuclei (AGNs) in a statistically significant sample of bulgeless 
(Sd/Sdm) disk galaxies. Based on the detection of the high-ionization [NeV] 
14.3~$\mu$m line, we report the discovery of an AGN in one out of the 18 
galaxies in the sample. This galaxy, NGC 4178, is a nearby edge-on Sd galaxy, 
which likely hosts a prominent nuclear star cluster (NSC). The bolometric 
luminosity of the AGN inferred from the [NeV] line luminosity is $\sim$ 
8$\times$10$^{41}$ ergs s$^{-1}$. This is almost two orders of magnitude 
greater than the luminosity of the AGN in NGC 4395, the best studied AGN 
in a bulgeless disk galaxy.  Assuming that the AGN in NGC 4178 is radiating 
below the Eddington limit, the lower mass limit for the black hole is  
$\sim$ 6$\times$10$^3$M$_{\odot}$.  The fact that none of the other galaxies 
in the sample shows any evidence for an AGN demonstrates that while the AGN 
detection rate based on mid-infrared diagnostics is high (30-40\%) in 
optically quiescent galaxies with pseudobulges or weak classical bulges 
(Hubble type Sbc and Sc), it drops drastically in Sd/Sdm galaxies. Our 
observations therefore confirm that AGNs in completely bulgeless disk 
galaxies are {\it not} hidden in the optical but truly are rare.  Of the 
three Sd galaxies with AGNs known so far, all have prominent NSCs, 
suggesting that in the absence of a well-defined bulge, the galaxy must 
possess a NSC in order to host an AGN. On the other hand, while the presence 
of a NSC appears to be a requirement for hosting an AGN in bulgeless galaxies,
 neither the properties of the NSC nor those of the host galaxy appear 
exceptional in late-type AGN host galaxies. The recipe for forming and 
growing a central black hole in a bulgeless galaxy therefore remains unknown.

\end{abstract}


\section{Introduction}

We now know that supermassive black holes lurk in the centers of most 
bulge-dominated galaxies in the local Universe and that their black hole 
masses, M$_{\rm BH}$, and the stellar velocity dispersions, $\sigma$, of 
their host galaxies are strongly correlated (Gebhardt et al. 2000, 
Ferrarese \& Merritt 2000). This discovery has launched numerous speculations 
that the formation and evolution of galaxies and supermassive black holes are 
fundamentally linked, and that perhaps the presence of a bulge is a necessary 
ingredient for a black hole to form and grow. Indeed, M33, the most nearby 
example of a truly bulgeless disk galaxy shows no evidence of a supermassive 
black hole, and the upper limit on the black hole mass determined by stellar 
dynamical studies is significantly below that predicted by the 
M$_{\rm BH}$-$\sigma$  relation established in early-type galaxies (e.g., 
Gebhardt et al. 2001). In contrast, the disk galaxy NGC 4395 shows no 
evidence for a bulge and yet does contain an active nucleus (e.g., 
Filippenko \& Ho 2003). However, this galaxy has remained until recently the 
only case of a truly bulgeless disk galaxy with an accreting black hole, 
leaving open the possibility that it is an anomaly. Indeed, prior to the 
launch of {\it Spitzer}, the vast majority of known accreting black holes - 
i.e., active galactic nuclei (AGN) - in the local Universe were found in 
galaxies with prominent bulges (e.g. Heckman 1980; Ho, Filippenko, \& 
Sargent 1997; Kauffmann et al. 2003). 

However, these studies were based on spectroscopic observations at optical 
wavelengths, which can be severely limited in the study of bulgeless galaxies,
 where a putative AGN is likely to be both energetically weak and deeply 
embedded in the center of a dusty late-type spiral. In such systems, the 
traditional optical emission lines used to identify AGN can be dominated by 
emission from star formation regions, in addition to being significantly 
attenuated by dust in the host galaxy. As a result, it is by no means clear 
what fraction of late-type galaxies host AGN. Therefore, some key fundamental 
questions on the connection between black holes and galaxy formation and 
evolution have yet to be answered, such as: What fraction of late-type 
galaxies host AGNs? Do black holes form and grow in galaxies without a bulge? 
How are the incidence and properties of the black hole related to the host 
galaxy in cases where there is no bulge?

Motivated by these questions, and by the possibility that optical studies 
may fail at finding AGNs in the latest Hubble types, we have previously 
conducted an exhaustive archival mid-infrared (MIR) spectroscopic 
investigation of 34 late-type (Sbc or later) galaxies observed by 
{\it Spitzer} to search for AGNs (Satyapal et al. 2007; Satyapal et al. 2008 
- henceforth S07; S08, respectively). 
Remarkably, these observations revealed the presence of the high ionization 
[NeV] 14~\micron\ and/or 24~\micron\ lines - which are not generally produced 
in ionized gas surrounding hot stars - in a significant number of galaxies 
that have no clear signatures of an AGN in their optical spectra. Using 
detailed photoionization models with both an input AGN and an extreme 
EUV-bright radiation field from a young starburst, we demonstrated that the 
MIR spectrum of these galaxies cannot be replicated unless an AGN 
contribution, in some cases as weak as 10\% of the total galaxy luminosity, 
is included (S08, Abel \& Satyapal 2007). {\it This implies that the AGN 
detection rate in late-type galaxies is possibly more than 4 times larger 
than what optical spectroscopic studies alone indicate}. We have obtained 
follow-up X-ray observations of a subset of these galaxies, which in all 
cases confirm the presence of an AGN (Gliozzi et al. 2009; Satyapal et al. 
2009). A more recent {\it Spitzer} study has also uncovered a significant 
population of AGNs in optically quiescent galaxies of earlier Hubble type 
(Goulding \& Alexander 2009), demonstrating the power of mid-infrared 
spectroscopy in AGN searches. Other recent multiwavelength studies have also 
shown that AGNs are significantly more common in late-type galaxies than 
once thought (e.g., Greene, Ho, \& Barth 2009; Shields et al. 2008; Ghosh 
et al. 2008; Barth et al. 2009; Dewangan et al. 2008; Desroches \& Ho 2009). 
It is therefore clear that classical bulges are {\it not} required for black 
holes to form and grow.

While it is evident that AGNs do reside in a significant number of late-type 
galaxies, most of the galaxy hosts appear to have a pseudobulge component, 
i.e. a central light excess 
characterized by an exponential surface brightness profile. These 
pseudobulges are thought to 
form via secular processes, in contrast to the violent merger-driven 
formation history of classical bulges (Kormendy\& Kennicutt 2004). Amongst 
our previous archival Spitzer sample, there were only 4 very late-type 
spirals (Hubble type Sd/Sdm) without any obvious sign of a pseudo-bulge. 
We discovered prominent [NeV] emission from only one of these sources - 
the nearby Sd galaxy NGC 3621 (S07). Follow-up X-ray (Gliozzi et al. 2009) 
and high spatial resolution optical (Barth et al. 2009) observations confirm 
the presence of an AGN in this source. NGC 3621 is similar to NGC 4395 in 
that both galaxies are essentially bulgeless and contain a massive nuclear 
star cluster (Barth et al. 2009). However, with only two examples and only a 
total of 4 truly bulgeless disk galaxies observed, it is not possible to 
determine robustly the fraction of AGNs in pure disk galaxies and to 
understand how the incidence and properties of black holes relate to the host 
galaxy in the absence of a bulge.

In this paper, we present results from a recent {\it Spitzer} MIR 
spectroscopic investigation of 18 optically quiescent, truly bulgeless disk 
galaxies in order to search for previously undetected low luminosity and/or 
embedded AGN. This is the first systematic MIR search for weak or hidden AGN 
in a statistically significant sample of ''pure`` disk galaxies. The primary 
goal of this paper is to refine the incidence of AGNs in this type of galaxy.
  As our previous {\it Spitzer} work has demonstrated, optical studies miss 
a significant fraction of AGNs in late-type galaxies, leaving open the 
possibility that there are a significant number of active black holes in the 
centers of completely bulgeless galaxies that are as yet undiscovered.

This paper is structured as follows.  In Section 2, we summarize the 
properties of the {\it Spitzer} sample presented in this paper.  In Section 3,
 we summarize the observational details and data analysis procedure, followed 
by a description of our results in Section 4.  In Section 5, we discuss the 
origin of the [NeV] emission and the evidence for an AGN in our sample, 
followed by a discussion of the AGN detection rate in pure disk galaxies in 
Section 6. In Section 7, we investigate the demographics of late-type 
galaxies with AGNs, followed by an exploration in Section 8 of the host 
characteristics of the few AGNs that reside in definitively bulgeless 
galaxies. A summary of our major conclusions is given in Section 9.

\section{The Sample}

Our goals in selecting a sample were to 1) obtain a statistically 
significant sample of pure (i.e. bulgeless) disk galaxies to provide 
meaningful estimates of the fraction that host AGNs, 2) select close-by 
objects to enable detailed follow-up of potential AGN discoveries, 3) 
select isolated disk galaxies to avoid the effects of interactions on 
triggering black hole formation and growth and to test whether black holes 
routinely form through purely secular processes, and 4) select well-studied 
sources with extensive optical spectroscopic and multiwavelength data 
available in the literature. Our target sources were selected from the 
Palomar survey of nearby bright galaxies (Ho et al. 1997; henceforth H97).  
Of the 486 galaxies in the Palomar survey, a little more than two-dozen are 
of Hubble type of Sd/Sdm.  Excluding galaxies with irregular morphologies or 
other signs of interactions, and excluding the well-studied Seyfert NGC 4395, 
our final sample contained 18 galaxies.

Table 1 summarizes the basic properties of the galaxies in our sample.  All 
targets are nearby, ranging in distance from 2.4 to 21.6 Mpc, with an average 
distance of $\sim$ 11 Mpc. The aperture of the optical measurements from H97 
was 2''$\times$4''. The extinction-corrected absolute B-band magnitude ranges 
from $\sim$ -17 to -20. We estimate and list in Table 1 galaxy stellar masses 
using the extinction-corrected B-V color and absolute B magnitude using the 
mass-to-light ratios from Bell et al. (2003).  The inferred galaxy masses for 
our sample range from  $\sim$ 8$\times$10$^8$M$_{\odot}$ to $\sim$ 
1.6$\times$10$^{10}$M$_{\odot}$. The distribution of derived galaxy masses of 
the sample is shown in Figure 1.  We also list in Table 1 the nuclear star 
formation rate (SFR), estimated using the extinction-corrected H$\alpha$ 
luminosity from H97, assuming all of the H$\alpha$ luminosity arises from 
star forming regions, and using the prescription given in Kennicutt (1998).  
The SFR ranges from $\sim$ 6$\times$10$^{-5}$M$_{\odot}$/yr to $\sim$ 
2$\times$10$^{-3}$M$_{\odot}$/yr.  The (inclination-corrected) line width of 
the HI profile, also from H97, ranges from 21 to 314 km/s, suggesting that 
the sample spans a large range of dark matter mass. 
To determine whether the presence or properties of potential AGNs are related 
in any way to gas mass, we list in Table 1 the HI mass, estimated from the HI 
fluxes compiled in H97.

\begin{figure}[]

\noindent{\includegraphics[width=9cm]{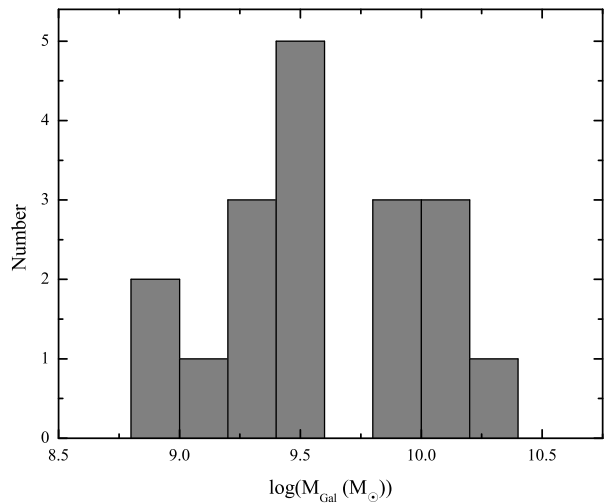}}
\caption[]{The distribution of galaxy masses for the sample.  Galaxy masses 
were estimated using the extinction-corrected B-V color and B-band magnitude 
taken directly from H97 using the M/L ratios from Bell et al. (2003)
}
\end{figure}

\begin{table*}
\fontsize{7pt}{7pt}\selectfont
\begin{center}
\begin{tabular}{lclcccccccccc}
\multicolumn{13}{c}{{\bf Table 1: Properties of the Sample}}\\ 
\hline
\hline
\noalign{\smallskip}
Galaxy & Distance & [OIII]/H$_{\beta}$ & [OI]/H$_\alpha$ & [NII]/H$_\alpha$ & [SII]/H$_\alpha$ & Optical & M$_{B_T}^0$ & M$_{Gal}$ & M$_{HI}$ & $\Delta{\rm V}_{rot}$ & NC & SFR \\
Name & (Mpc) & & & & &  Class & & (M$_{\odot}$) & M$_{\odot}$ & (km/s) & & (10$^{-3}{\rm M}_{\odot}$/yr)\\
\noalign{\smallskip}
\hline
\noalign{\smallskip}
IC 2574& 3.4& 0.23& 0.025& 0.07& 0.25& H& -17.33& 8.91& 9.05& 131& Unlikely & 0.16\\
NGC 2500& 10.1& 2.42& 0.068& 0.33& 0.45& H& -18.03& 9.40& 8.91& 228& Yes & 0.14\\
NGC 2537& 9& 1.83& 0.008& 0.15& 0.18& H& -17.75& 9.29& 8.59& 205& $\cdots$ & 27.81\\
NGC 3027& 19.5& 1.16& 0.037& 0.19& 0.53& H& -19.77& 10.10& 9.93& 247& $\cdots$ & 3.99\\
NGC 3432& 7.8& 1.71& 0.012& 0.14& 0.22& H& -18.43& 9.56& 9.26& 260& Maybe & 5.47\\
NGC 3495& 12.8& 0.38& 0.05& 0.42& 0.43& H& -19.8& 10.11& 9.17& 314& $\cdots$ & 0.67\\
NGC 4145& 20.7& 1.09& 0.13& 0.61& 0.8& T2& -20.04& 10.21& 9.82& 21& $\cdots$ & 1.68\\
NGC 4178& 16.8& 0.35& 0.022& 0.32& 0.5& H& -19.78& 10.10& 9.66& 299& Yes & 5.16\\
NGC 4242& 7.5& 1& 0.27& 0.27& 0.91& H& -18.08& 9.42& 8.78& 206& Yes & 0.06\\
NGC 4618& 7.3& 1.82& 0.012& 0.16& 0.23& H& -18.19& 9.47& 9.04& 220& Yes & 4.75\\
NGC 4656& 7.2& 4.02& 0.022& 0.05& 0.18& H& -19.19& 9.87& 9.62& 179& Unlikely & 2.05\\
NGC 4713& 17.9& 1.01& 0.087& 0.44& 0.67& T2& -19.41& 9.96& 9.64& 242& Yes & $<$ 3.14\\
NGC 5147& 21.6& 0.44& 0.035& 0.37& 0.52& H& -19.38& 9.94& 9.30& 259& $\cdots$ & 4.14\\
NGC 5204& 4.8& 0.96& 0.099& 0.18& 0.56& H& -16.93& 8.96& 8.76& 154& Maybe & $<$ 0.38\\
NGC 5585& 7& 1.7& 0.017& 0.15& 0.32& H& -18.18& 9.46& 9.19& 200& Yes & 2.63\\
NGC 6689& 12.2& 1.87& 0.067& 0.4& 0.62& H& -18.48& 9.58& 9.13& 220& Unlikely & 0.79\\
NGC 784& 4.7& 4.9& 0.006& 0.03& 0.09& H& -17.2& 9.07& 8.54& 116& Unlikely & $<$ 0.96\\
NGC 959& 10.1& 1.16& 0.032& 0.34& 0.37& H& -17.66& 9.26& 8.45& 196& Yes & 2.09\\
\hline
\end{tabular}
\end{center}
{\scriptsize{\bf Columns Explanation:} Col(1): Common Source Names; 
Col(2):  Distance to the source in units of Mpc are all taken directly 
from H97 where distances are adopted from Tully \& Shaya (1984);
Col(3): [OIII] to H$_\beta$ ratio taken from H97;
Col(4): [OI] to H$_\alpha$ ratio taken from H97;
Col (5): [NII] to H$_\alpha$ ratio taken from H97);
Col(6):  [SII] to H$_\alpha$ ratio taken from H97);
Col(7):  Optical classification of the source; ``H'' signifies HII region 
ratios, ``T'' represents transitional spectra between LINERs and HII 
regions, and ``2'' indicates that broad permitted lines were not found in 
the optical spectrum.
Col(8):  Total absolute B magnitude corrected for extinction, adopted from 
H97;
Col(9): Galactic Mass obtained from B-V color and B-magnitude from H97 
using mass-to-light ratios from Bell et al. (2003);  
Col(10): HI mass taken from H97;
Col(11): Inclination-corrected Hi rotational amplitude taken directly from 
Table 10, col(7) in H97;
Col(12) Presence of Nuclear Cluster based on archival {\it HST} observations.
Col (13) Nuclear star formation rate estimated using the 
extinction-corrected H$\alpha$ luminosity from H97, assuming all of the 
H$\alpha$ luminosity arises from star forming regions, and using the 
prescription given in Kennicutt (1998)}
\end{table*}

 The majority of galaxies in our sample are classified in H97 as ``HII'' 
stellar-powered galaxies; only two are ``T2'' transition galaxies.  T2 
galaxies have optical line ratios intermediate between HII galaxies and 
low-ionization nuclear emission-line regions (LINERs) and have no broad 
permitted lines (e.g. H$\alpha$) in their optical spectrum.  There is 
therefore no firm optical spectroscopic evidence for AGNs in any of the 
galaxies in our sample. This is illustrated in Figure 2 which shows the 
standard optical line ratio diagnostic diagrams (Veilleux \& Osterbrock 
1987) widely used to classify AGNs for the entire H97 sample. Our Spitzer 
sample is highlighted, together with the theoretical starburst limit line 
from Kewley et al. 2001, i.e. the maximum line ratios allowed by starburst 
photoionization models using the hardest possible radiation field.  Note 
that the majority of galaxies in our sample have optical line ratios well 
to the left of this line, indicating that the optical line ratios do not 
require the presence of {\it any} AGN contribution.

\begin{figure*}[htbp]
\begin{center}
\begin{tabular}{cc}
  \includegraphics[width=0.45\textwidth]{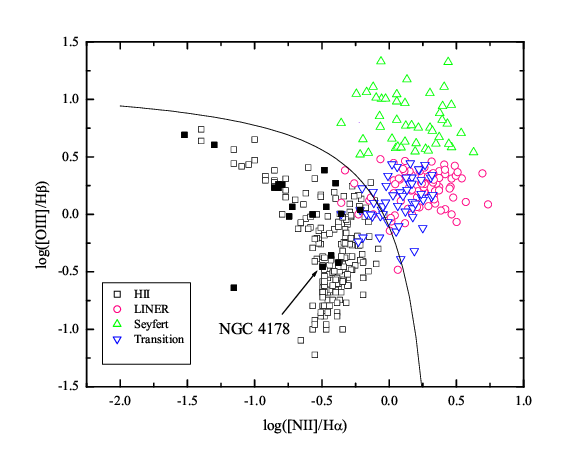} &
  \includegraphics[width=0.45\textwidth]{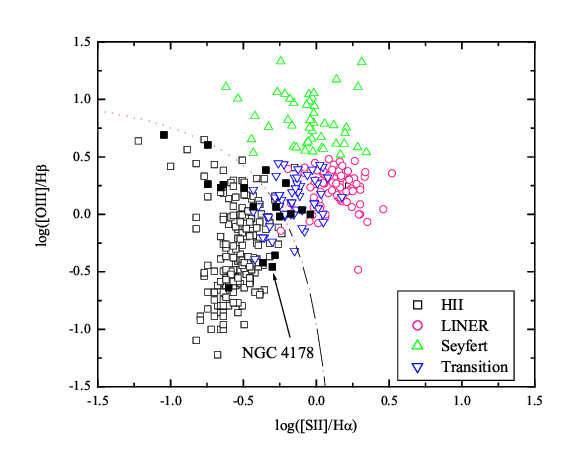} \\
  \multicolumn{2}{c}{\includegraphics[width=0.45\textwidth]{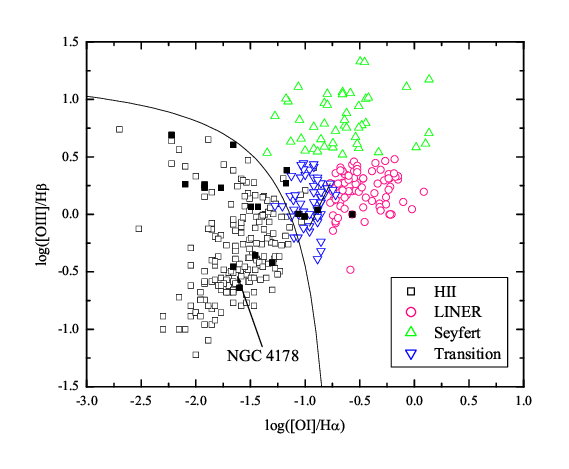}} \\
\end{tabular}
\end{center}
\caption[]{Standard optical line ratio diagnostic diagrams (Veilleux \& 
Osterbrock 1987) for the entire H97 sample with our current {\it Spitzer} 
sample highlighted by filled symbols. The solid line indicates the 
theoretical starburst limit from Kewley et al. 2001.  This limit marks the 
maximum line ratios possible from starburst-only photoionization models.  
NGC 4178, the only galaxy in our sample with a [NeV] detection, is labeled 
in the figure. The optical line ratios for this galaxy are well to the 
left of the starburst limit, indicating that the optical line ratios are 
consistent with ionization by star formation only.}
\end{figure*}

As can be seen from the DSS images in Figure 3, the morphologies of the 
sample galaxies are varied, with some objects having prominent disks and 
clear photocenters, and others displaying a more diffuse structure with no 
obvious photocenter. 
Archival {\it HST} images are available for 14 out of the 18 galaxies in 
the sample, taken mostly
with the {\it Advanced Camera for Surveys} (ACS).  We inspected these 
images for the presence
of a well-defined nuclear star cluster (NSC). An unambiguous detection of 
such a source was possible in 7 of the 14 sample galaxies. These galaxies 
are identified in Table 1. 
We also performed elliptical isophote fitting to check for a notable bulge 
component in the one-dimensional surface brightness profiles, and 
confirmed that all of the sample galaxies are
pure disk galaxies.

\begin{figure*}[htbp]
\begin{center}
\begin{tabular}{ccc}
  \includegraphics[width=0.3\textwidth]{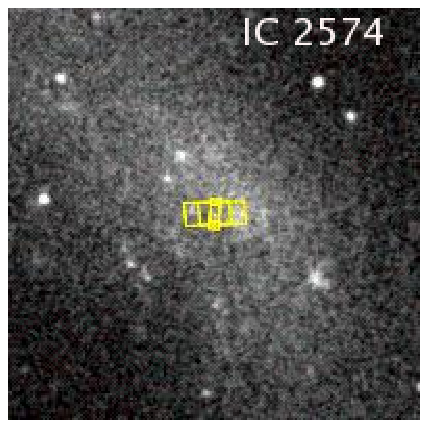} &
  \includegraphics[width=0.3\textwidth]{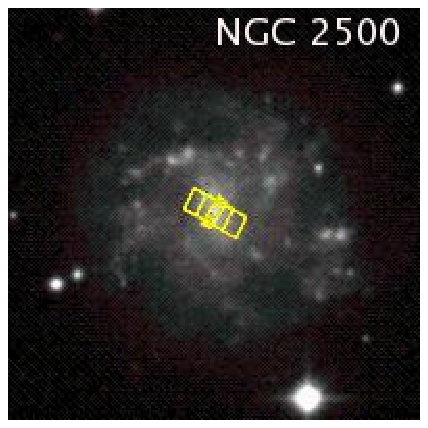} &
  \includegraphics[width=0.3\textwidth]{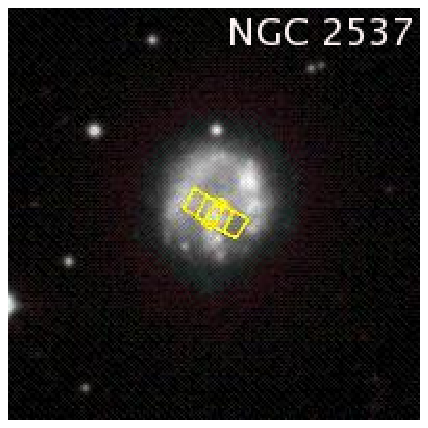} \\
  \noalign{\smallskip}
\noalign{\smallskip}

  \includegraphics[width=0.3\textwidth]{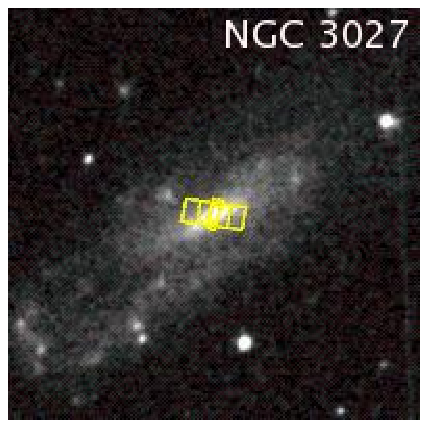} &
  \includegraphics[width=0.3\textwidth]{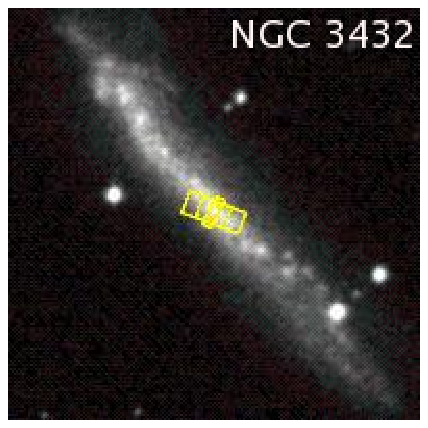} &
  \includegraphics[width=0.3\textwidth]{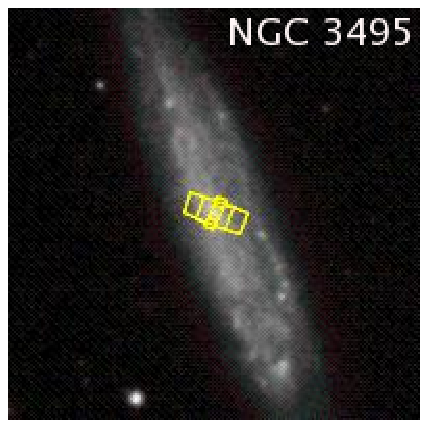} \\
  \noalign{\smallskip}
\noalign{\smallskip}

\includegraphics[width=0.3\textwidth]{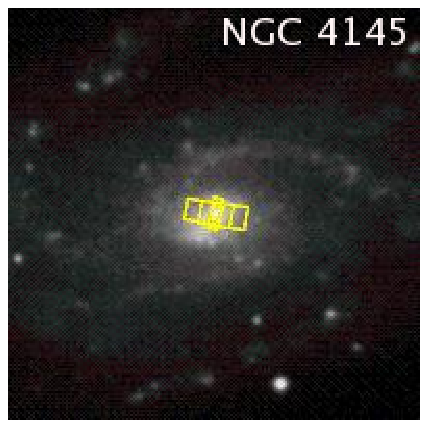} &
  \includegraphics[width=0.3\textwidth]{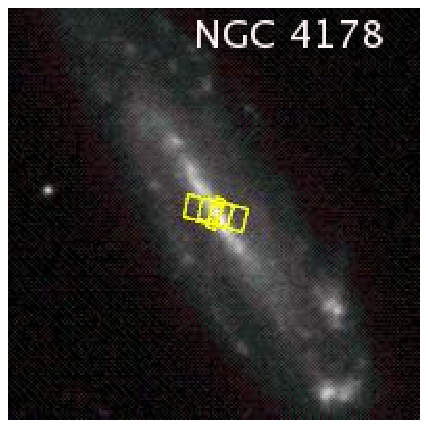} &
  \includegraphics[width=0.3\textwidth]{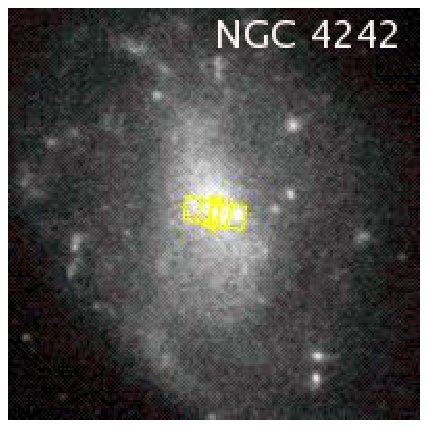} \\

\end{tabular}
\end{center}
\caption[]{DSS images of the galaxies in the sample with IRS SH and LH 
slits overlayed. The field of view of the images is 200"$\times$200".
}
\end{figure*}

\begin{figure*}[htbp]
\begin{center}
\begin{tabular}{ccc}
  \includegraphics[width=0.3\textwidth]{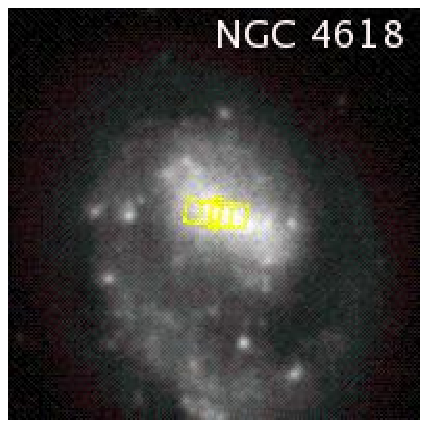} &
  \includegraphics[width=0.3\textwidth]{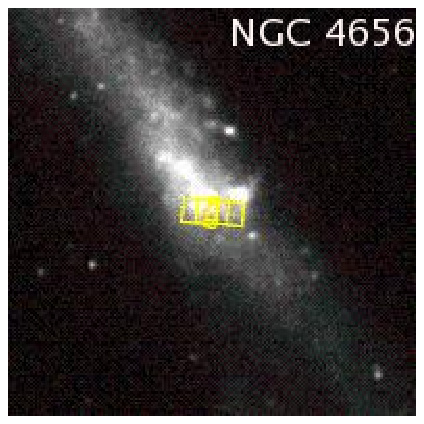} &
  \includegraphics[width=0.3\textwidth]{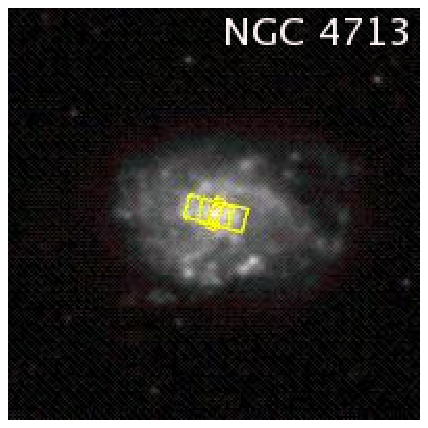} \\
  \noalign{\smallskip}
\noalign{\smallskip}

  \includegraphics[width=0.3\textwidth]{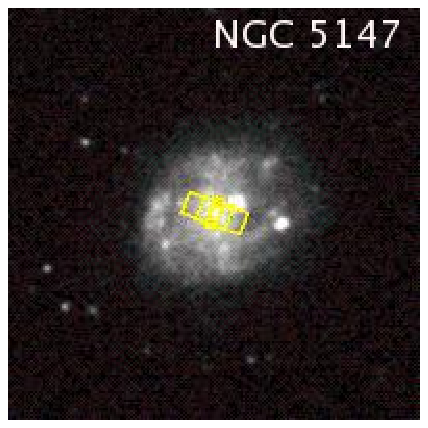} &
  \includegraphics[width=0.3\textwidth]{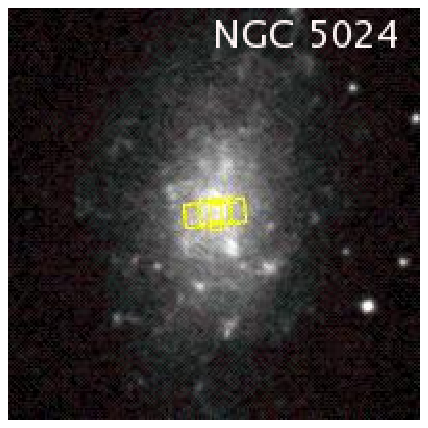} &
  \includegraphics[width=0.3\textwidth]{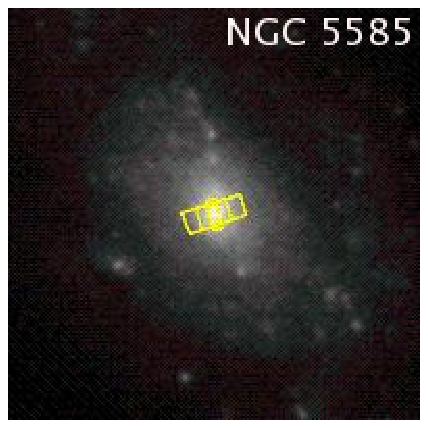} \\
  \noalign{\smallskip}
\noalign{\smallskip}

\includegraphics[width=0.3\textwidth]{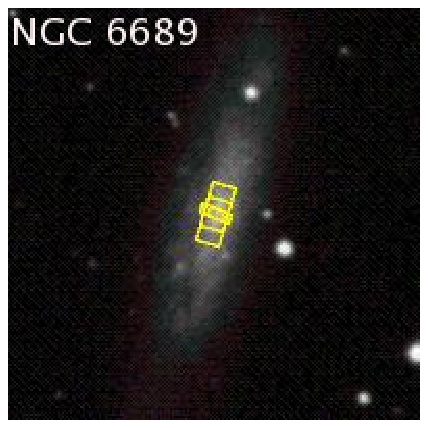} &
  \includegraphics[width=0.3\textwidth]{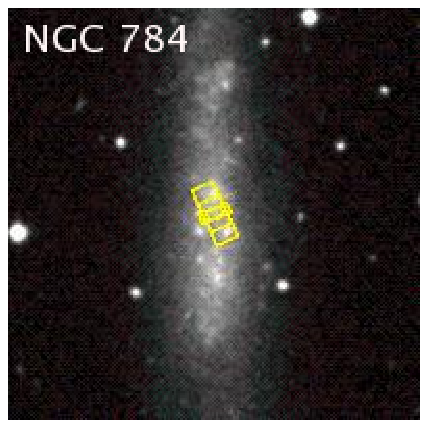} &
  \includegraphics[width=0.3\textwidth]{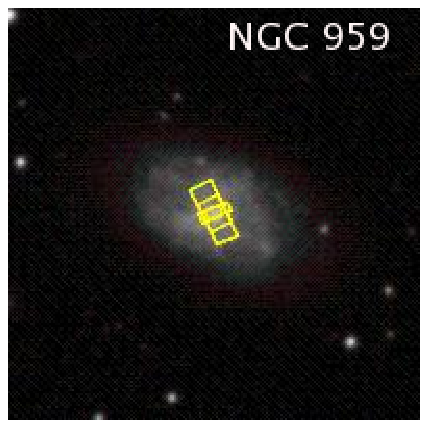} \\

  \multicolumn{3}{c}{\sc{Fig} 3.--{\it Continued}}\\
\end{tabular}
\end{center}
\end{figure*}

\section{Observations and Data Reduction}

Data for all but one of the galaxies in our sample were obtained from the 
{\it Spitzer} Cycle 5 GO program ID 50339 (PI: Satyapal).  These 
observations were all executed between 2008 June and 2008 November.  
One of the galaxies in our sample (NGC 4618) was previously observed 
with the IRS in 2005 December (PID 20140). We used the archival data for 
this galaxy.  All observations were carried out in staring mode using both 
the short-wavelength (SH, 4.7''$\times$11.3'', $\lambda$  = 9.9-19.6~$\mu$m)
 and long-wavelength (LH, 11.1''$\times$22.3'', 
$\lambda$ = 18.7-37.2~$\mu$m) high-resolution modules of the Infrared 
Spectrograph (IRS; Houck et al. 2004) which have a spectral resolution of 
R $\sim$ 600. Exposure times were chosen to provide a signal-to-noise ratio 
of at least 5 for the [NeV] 14.3~~$\mu$m line, assuming the lowest [NeV] 
luminosity detected to date in any galaxy (S08). 
Each observation was followed by a background sky observations located 
2~$\arcmin$ from the source in order to enable backround-subtraction. 

All on-source observations were centered on the galaxy nuclei, i.e. the 
photocenter coordinates 
from H97, which agree well with the 2MASS coordinates. Figure 3 overlays 
the SH and LH slit apertures onto the DSS images for all our targets, 
demonstrating that the nucleus of the galaxy always falls well within the 
slit. The slit size for the median distance of 10 Mpc corresponds to a 
projected extraction aperture of 0.2 kpc$\times$0.5 kpc and 0.5 
kpc$\times$1.1 kpc for the SH and LH modules, respectively. The projected 
slit sizes, as well as a number of other observational details, are listed 
in Table 2.

We note that the SH and LH staring observations are dithered, i.e. the 
integration is split into two slit positions overlapping by one third of 
a slit. Unless the emission originates from a compact source that falls 
entirely within the slit for both pointings, the two spectra cannot be 
averaged.  The procedure for flux extraction for staring observations was 
the following: 1) If the fluxes measured from the two slits differed by no 
more than the calibration error of the instrument, then the fluxes 
were averaged; otherwise, the slit with the highest measured line flux was 
chosen.  2) If an emission line was detected in one slit, but not in the 
other, then the detection was selected. The overall calibration uncertainty 
for the fluxes we report in this paper is 15\%.

The raw data were preprocessed by the IRS pipeline (version 17.2) at the 
{\it Spitzer} Science Center (SSC) prior to download.  Preprocessing 
includes ramp fitting, dark-sky subtraction, droop correction, linearity 
correction, flat-fielding, and flux 
calibration\footnote[2]{See {\it Spitzer} Observers Manual, Chapter 7, 
http://ssc.spitzer.caltech.edu/documents/som/som8.0.irs.pdf}. The 
{\it Spitzer} spectra were further processed using the SMART v. 6.3.0 
analysis package (Higdon et al. 2004) and the corresponding version of 
the calibration files (v.1.5.0), which were used to obtain final line 
fluxes.  Each spectra was individually inspected and any bad pixels 
remaining after pipeline processing were removed.  The fine-structure line 
fluxes presented in this work were obtained from Gaussian fits to the 
spectral line and linear fits to the baseline continuum.  

Many of the galaxies in our sample display prominent PAH features and 
molecular hydrogen emission lines.  In this work, we limit the discussion 
to the fine-structure emission lines relevant for identification of 
potential AGN. We defer discussion of all other spectral diagnostics and the 
star-formation properties of bulgeless galaxies to a future paper.

\begin{table*}
\fontsize{8pt}{8pt}\selectfont
\begin{center}
\begin{tabular}{lccrlcc}
\multicolumn{7}{c}{{\bf Table 2: Observational Details}}\\
\noalign{\smallskip}
\hline
\hline
\multicolumn{7}{l}{}\\
Galaxy & Exposure Time & Exposure Time & 
\multicolumn{2}{c}{Position (J2000)} & Extraction Aperture & {Extraction Aperture}\\

Name & SH (seconds) & LH (seconds) & \multicolumn{1}{c}{RA} & \multicolumn{1}{c}{Dec} & Size SH (pc) & Size LH (pc) \\
\noalign{\smallskip}

\hline
\multicolumn{7}{l}{}\\

IC2574 & 6$\times$3 & 6$\times$3 & 10 28 23.50 & +68 24 44.00 & 77$\times$186 & 183$\times$368\\
NGC2500 & 30$\times$3 & 14$\times$3 & 08 01 53.21 & +50 44 13.60 & 230$\times$553 & 544$\times$1092\\
NGC2537 & 30$\times$2 & 14$\times$3 & 08 13 14.60 & +45 59 23.30 & 205$\times$493 & 484$\times$973\\
NGC3027 & 120$\times$3 & 240$\times$3 & 09 55 40.60 & +72 12 12.80 & 444$\times$1068 & 1049$\times$2108\\
NGC3432 & 30$\times$2 & 14$\times$2 & 10 52 31.13 & +36 37 07.60 & 178$\times$427 & 420$\times$843\\
NGC3495 & 30$\times$4 & 60$\times$3 & 11 01 16.23 & +03 37 40.50 & 292$\times$701 & 689$\times$1384\\
NGC4145 & 120$\times$4 & 240$\times$4 & 12 10 01.52 & +39 53 01.90 & 472$\times$1134 & 1114$\times$2238\\
NGC4178 & 120 $\times$2 & 240$\times$2 & 12 12 46.40 & +10 51 57.50 & 383$\times$920 & 904$\times$1816\\
NGC4242 & 30$\times$2 & 14$\times$2 & 12 17 30.18 & +45 37 09.50 & 171$\times$411 & 404$\times$811\\
NGC4618 & 30$\times$2 & 60$\times$2 & 12 11 32.85 & +41 09 02.80 & 166$\times$400 & 393$\times$789\\
NGC4656 & 30$\times$2 & 14$\times$2 & 12 43 57.73 & +32 10 05.30 & 164$\times$394 & 387$\times$778\\
NGC4713 & 120$\times$2 & 240$\times$2 & 12 49 57.87 & +05 18 41.10 & 408$\times$981 & 963$\times$1935\\
NGC5147 & 120$\times$4 & 240$\times$5 & 13 26 19.71 & +02 06 02.60 & 492$\times$1183 & 1162$\times$2335\\
NGC5204 & 6$\times$3 & 6$\times$3 & 13 29 36.51 & +58 25 07.40 & 109$\times$263 & 258$\times$519\\
NGC5585 & 30$\times$2 & 14$\times$2 & 14 19 48.20 & +56 43 44.60 & 160$\times$383 & 377$\times$757\\
NGC6689 & 30$\times$3 & 60$\times$2 & 18 34 50.25 & +70 31 26.10 & 278$\times$668 & 657$\times$1319\\
NGC784 & 6$\times$3 & 6$\times$3 & 02 01 16.93 & +28 50 14.10 & 107$\times$257 & 253$\times$508\\
NGC959 & 30$\times$2 & 14$\times$3 & 02 32 23.94 & +35 29 40.70 & 230$\times$553 & 544$\times$1092\\ 
\multicolumn{7}{l}{}\\
\hline
\end{tabular}
\end{center}
{\scriptsize{\bf Columns Explanation:} 
Col(1):  Common Source Names; 
Col(2) \& (3):  On-source exposure time per pointing in seconds given for 
the SH and LH modules, respectively;
Col(4) \& (5):  Coordinates used for each observations;
Col(6) \& (7):  Extraction apertures for the SH and LH modules, which 
correspond to the full aperture of the slit.  Values given are in parsecs, 
using galaxy distances listed in Table 1.}
\end{table*}

\section{Results}

\subsection{ Fine-Structure Line Fluxes }

In Table 3 we list the line fluxes, statistical errors and 3$\sigma$ upper 
limits for the strongest MIR fine-structure lines.  The apertures from which 
these fluxes were extracted are listed in Table 2.  In all cases, detections 
were defined when the line flux was at least 3$\sigma$.  The strongest 
emission features in the spectra were the [NeII] 12.8~$\mu$m, [NeIII] 
15.5~$\mu$m, [SIII] 18.7~$\mu$m, [SIII] 33.5~$\mu$m, and [SiIII] 34.8~$\mu$m 
lines, detected in $\sim$ 60\%-80\% of the galaxies in the sample.  The 
[FeII] 26~$\mu$m, [NeV] 24~$\mu$m, and [OIV] 25.9~$\mu$m emission lines were 
not detected in any galaxy in the sample. We note that the spectral 
resolution of the SH and LH modules of the IRS is insufficient to resolve 
the velocity structure for most of the lines.  We detected the [NeV] 
14.3$\mu$m line in only 1 out of the 18 galaxies, providing strong evidence 
for the presence of an AGN in this one galaxy.  We discuss the IR spectral 
line fluxes and flux ratios for this galaxy, NGC 4178, separately in Section 
5 below.
\clearpage
\begin{sidewaystable}
\fontsize{7pt}{7pt}\selectfont
\begin{center}
\begin{tabular}{lrrrrrrrrrr}
\multicolumn{11}{c}{{\bf Table 3: Fine-Structure Line Fluxes}}\\ 
\hline
\hline
\multicolumn{11}{l}{}\\
Name & [SIV] 10.51\micron & [NeII] 12.81\micron & [NeV] 14.32\micron & [NeIII] 15.56\micron & [SIII] 18.71\micron & [NeV] 24.32\micron & [OIV] 25.89\micron & [FeII] 25.99\micron & [SIII] 33.48\micron & [SiII] 34.82\micron\\
\hline
\multicolumn{11}{l}{}\\
IC2574 & $<$19.95 & $<$10.25 & $<$19.31 & $<$13.18 & $<$30.29 & $<$22.96 & $<$24.47 & $<$24.56 & $<$89.63 & $<$119.89\\
NGC2500 & $<$6.46 & $<$3.87 & $<$5 & $<$3.69 & $<$5.34 & $<$9.69 & $<$10.32 & $<$10.36 & 90$\pm$19 & $<$27.51\\
NGC2537 & $<$5.88 & 9.6$\pm$2.76 & $<$3.4 & $<$3.95 & $<$4.15 & $<$8.16 & $<$8.69 & $<$8.72 & 43$\pm$13 & $<$43.89\\
NGC3027 & 7.6$\pm$1.02 & 6.13$\pm$0.62 & $<$0.48 & 3.1$\pm$0.6 & 4.81$\pm$0.55 & $<$2.62 & $<$2.8 & $<$2.81 & 21.5$\pm$2.88 & 21.2$\pm$2.46\\
NGC3432 & 8.1$\pm$1.75 & 28.1$\pm$2.5 & $<$3.71 & 30.2$\pm$3.4 & 28.9$\pm$2.16 & $<$7.75 & $<$8.25 & $<$8.29 & 198.5$\pm$15.16 & 122.5$\pm$8.59\\
NGC3495 & $<$2.95 & 13.3$\pm$1.3 & $<$1.88 & 12.2$\pm$1.6 & 14$\pm$3.8 & $<$5.45 & $<$5.81 & $<$5.83 & 32.9$\pm$9 & 39.7$\pm$7.5\\
NGC4145 & 5.13$\pm$1.42 & 7.55$\pm$0.87 & $<$0.64 & 2.77$\pm$0.36 & 3.65$\pm$0.55 & $<$1.89 & $<$2.02 & $<$2.03 & 17$\pm$7.19 & 22.8$\pm$0.65\\
NGC4178 & $<$1.23 & 20.9$\pm$1.95 & 2.43$\pm$0.39 & 5.95$\pm$0.45 & 12.82$\pm$0.89 & $<$2.79 & $<$2.11 & $<$2.14 & 80.7$\pm$4.6 & 83.7$\pm$2.53\\
NGC4242 & $<$2.83 & $<$3.78 & $<$3.47 & $<$4.58 & $<$5.26 & $<$9.44 & $<$10.05 & $<$10.09 & $<$26.44 & $<$31.15\\
NGC4496 & $<$5.49 & $<$6.53 & $<$3.58 & $<$4.5 & $<$6.6 & $<$5.44 & $<$5.8 & $<$5.82 & $<$9.19 & $<$15.38\\
NGC4618 & $<$8.24 & 36.59$\pm$6.62 & $<$4.78 & 24.91$\pm$3.88 & 24.51$\pm$1.94 & $<$8.84 & $<$9.6 & $<$9.64 & 63.23$\pm$6.82 & 49.41$\pm$6.25\\
NGC4656 & 47.3$\pm$5.25 & 40$\pm$1.7 & $<$3.5 & 41.8$\pm$2.65 & 23.8$\pm$3.7 & $<$6.82 & $<$7.26 & $<$7.29 & 128$\pm$15.49 & 97.8$\pm$14\\
NGC4713 & $<$1.97 & 15.85$\pm$1.58 & $<$1.85 & 5.47$\pm$0.92 & 7.96$\pm$1.07 & $<$2.21 & $<$2.36 & $<$2.37 & 92.2$\pm$4.18 & 94.9$\pm$2.63\\
NGC5147 & $<$1.46 & 9.06$\pm$1.09 & $<$1 & 3.6$\pm$0.42 & 4.96$\pm$0.7 & $<$2.3 & $<$2.45 & $<$2.46 & 28.8$\pm$3.47 & 32.8$\pm$2.13\\
NGC5204 & $<$16.99 & $<$8.06 & $<$9.01 & $<$11.16 & 51.4$\pm$7.25 & $<$10.57 & $<$11.26 & $<$11.31 & 165$\pm$54.3 & $<$78.34\\
NGC5585 & $<$2.52 & 11.6$\pm$2.9 & $<$2.04 & 11.7$\pm$2.15 & 17.1$\pm$2.74 & $<$7.48 & $<$7.97 & $<$8 & 53.5$\pm$14 & 67.8$\pm$16.3\\
NGC6689 & $<$3.68 & $<$3.32 & $<$3.14 & $<$2.68 & $<$4.96 & $<$5.94 & $<$6.33 & $<$6.36 & 36.35$\pm$9.59 & 32$\pm$3.8\\
NGC784 & $<$20.97 & $<$10.25 & $<$17.93 & 45$\pm$18.3 & $<$26.21 & $<$16.4 & $<$17.48 & $<$17.54 & $<$34.07 & $<$82.68\\
NGC959 & $<$5.15 & $<$3.42 & $<$3.69 & 9.25$\pm$1.9 & $<$5.92 & $<$7.7 & $<$8.21 & $<$8.24 & 60.3$\pm$12 & 47.7$\pm$14.2\\
\multicolumn{11}{l}{}\\
\hline
\end{tabular}
\end{center}
{\scriptsize{\bf Columns Explanation:} 
Col(1): Common Source Names; 
Col(2)-Col(11)):  Fluxes are in units of 10$^{-22}$ W cm$^{-2}$. 3 $\sigma$ upper limits are reported for nondetections.}
\end{sidewaystable}
\clearpage

\subsection{ Incidence of AGN }

The absence of [NeV] (ionization potential 97 eV) emission in our sample 
strongly suggests that with the exception of NGC 4178, none of the galaxies 
in our sample harbor AGNs.  In Table 4, we list the [NeV] 14.3~$\mu$m 
luminosities corresponding to the 3$\sigma$ upper limits on the fluxes for 
all galaxies in the sample with the exception of NGC 4178.  The luminosities 
were obtained using the galaxy distances listed in Table 1.  The upper 
limits to the line luminosity are well below 10$^{38}$ ergs s$^{-1}$. Using 
the compilations of MIR line fluxes of standard AGN from Sturm et al. (2002),
 Haas et al. (2005), Weedman et al. (2005), Ogle et al. (2006), Cleary et al.
 (2007), Armus et al. (2007), Gorjian et al. (2007), Deo et al. (2007), 
Tommasin et al. (2008), and Dale et al. (2009) \footnote[2]{Note that a few 
of the AGN were observed more than once.  In such cases, we always selected 
measurements that were obtained with the high resolution IRS module, 
choosing the reference with the largest compilation.}, there are 82 standard 
AGNs (optically classified as type 1 or type 2 AGN) with measured 
[NeV]~14$\mu$m line fluxes. The [NeV] 14~$\mu$m line luminosities for these 
AGNs range from $\sim$ 2$\times$10$^{38}$ ergs s$^{-1}$ to $\sim$ 
8$\times$10$^{42}$ ergs s$^{-1}$ with a median value of $\sim$ 
5$\times$10$^{40}$ ergs s$^{-1}$, more than two orders of magnitude above 
the [NeV] limiting sensitivities listed in Table 4.  The [NeV] luminosity 
of NGC 3621, our one and only previously discovered Sd galaxy with a weak 
AGN is $\sim$ 5$\times$10$^{37}$ ergs s$^{-1}$ (S07), consistent with or 
above the limiting sensitivities of $\sim$ 90\% of our sample.  Our 
non-detections thus firmly imply that these galaxies do not host AGNs with 
luminosities comparable to the weakest known in any galaxy. 

\begin{table*}
\fontsize{9pt}{9pt}\selectfont
\begin{center}
\begin{tabular}{lcccc}
\multicolumn{5}{c}{{\bf Table 4: [NeV] Luminosity Upper Limits and Line Flux Ratios}}\\
\noalign{\smallskip}
\hline
\hline
\multicolumn{5}{l}{}\\
Name &  $L_ {\rm [NeV]}$ &  ${\rm [OIV]_{25.89}/[NeII]_{12.81}}$  & ${\rm [NeV]_{14.32}/[NeII]_{12.81}}$ &
${\rm [SIII]_{18.71}/[SIII]_{33.48}}$ \\
\noalign{\smallskip}
\hline
\multicolumn{5}{l}{}\\
IC2574 & $<$2.52 &\nodata  &\nodata  &\nodata \\
NGC2500 & $<$5.77 &\nodata  &\nodata  &\nodata \\
NGC2537 & $<$3.12 & $<$0.91 & $<$0.35 &\nodata \\
NGC3027 & $<$2.06 & $<$0.46 & $<$0.08 & 1.67\\
NGC3432 & $<$2.55 & $<$0.29 & $<$0.13 & 1.91\\
NGC3495 & $<$3.48 & $<$0.44 & $<$0.14 & 1.56\\
NGC4145 & $<$3.09 & $<$0.27 & $<$0.08 & 0.51\\
NGC4242 & $<$2.21 &\nodata  &\nodata  &\nodata \\
NGC4496 & $<$6.94 &\nodata  &\nodata  &\nodata \\
NGC4618 & $<$2.88 & $<$0.26 & $<$0.13 & 3.59\\
NGC4656 & $<$2.05 & $<$0.18 & $<$0.09 & 1.54\\
NGC4713 & $<$6.72 & $<$0.15 & $<$0.12 & 1.90\\
NGC5147 & $<$5.28 & $<$0.27 & $<$0.11 & 1.43\\
NGC5204 & $<$2.35 &\nodata  &\nodata  & 0.95\\
NGC5585 & $<$1.13 & $<$0.69 & $<$0.18 & 1.22\\
NGC6689 & $<$5.28 &\nodata  &\nodata  &\nodata \\
NGC784 & $<$4.48 &\nodata  &\nodata  &\nodata \\
NGC959 & $<$4.25 &\nodata  &\nodata  &\nodata \\
\multicolumn{5}{l}{}\\
\hline
\end{tabular}
\end{center}
{\scriptsize{\bf Columns Explanation:} 
Col(1):  Common Source Names; 
Col(2):  [NeV] 14.32~\micron\ luminosity 3$\sigma$ upper limits in units 
of 10$^{37}~{\rm ergs ~s^{-1}}$;
Col(3) \& (4) \& (5):  Line flux ratios using fluxes from full 
apertures listed in Table 3.}

\end{table*}

There are a number of MIR diagnostics used to characterize the dominant 
ionizing radiation field in galaxies.  Since the flux ratio of emission 
lines from high-ionization to low-ionization ions depends on the nature of 
the ionizing source, the [NeV]~14.3~$\mu$m/[NeII]~12.8~$\mu$m and the  
[OIV]25.9$\mu$m/[NeII]12.8$\mu$m line flux ratios have been widely used to 
characterize the nature of the dominant ionizing source in galaxies 
(Genzel et al. 1996; Sturm et al. 2002; Satyapal et al. 2004; Dale et al. 
2006,2009).   We can compare our line flux ratio upper limits to the ratios 
in standard AGNs.  Again, using the recent compilations of MIR line fluxes 
of standard AGNs observed by {\it Spitzer} from Deo et al. (2007), Tommasin 
et al. (2008), and  Dale et al. (2009), there are 56 AGNs with measured 
[NeII] 12.8$\mu$m and [NeV]~14$\mu$m line fluxes.  The [NeV]/[NeII] line 
flux ratio in these galaxies ranges from 0.02 to 2.97, with a median value 
of 0.73. As can be seen from Table 4, all of the galaxies with 
[NeV]~14.3~$\mu$m upper limits have [NeV]/[NeII] upper limits well below the 
median value in standard AGNs, supporting the hypothesis that these galaxies 
lack an AGN.  Similarly, using the fluxes compiled in Verma et al. (2003), 
Deo et al. (2007), Tommasin et al. (2008), Dale et al. (2009), and Melendez 
et al. (2008), there are over 100 AGNs with measured [NeII] 12.8$\mu$m and 
[OIV]~25.9~$\mu$m line fluxes.  The [OIV]/[NeII] line flux ratio in these 
galaxies ranges from 0.02 to 11.1, with a median value of 1.33.  As can be 
seen from Table 4, the upper limits for the [OIV]/[NeII] flux ratio in our 
sample are also all well below the median value in standard AGNs, again 
strongly suggesting that these galaxies lack AGN.

An alternative diagnostic proposed by Dale et al. (2006) to classify the 
ionizing source in galaxies involves the [NeIII] 15.5$\mu$m/[NeII] 12.8 
$\mu$m flux ratio and the [SIII] 33.48$\mu$m/ [SiII]34.82$\mu$m flux ratio.  
They find that AGNs display lower [SIII] 33.48$\mu$m/[SiII] 34.82$\mu$m 
line flux ratios than "pure star-forming" nuclei, presumably due to enhanced 
[SiII] emission in X-ray dominated regions around AGNs (Maloney et al. 
1996).  In Figure 4 we plot the [NeIII] 15.5$\mu$m/[NeII] 12.8 $\mu$m flux 
ratio versus the [SIII] 33.48$\mu$m/[SiII]34.82$\mu$m flux ratio for 
standard AGN, starburst and HII galaxies (based on the compilations listed 
above), and for those galaxies in our sample for which the lines were 
detected.  We delineate the four regions defined by Dale et al. (2006), 
where regions I and II are exclusively occupied by LINERs and Seyferts, 
and regions III and IV are exclusively occupied by HII nuclei and 
extranuclear HII regions.  As can be seen, all of the galaxies in our 
sample (including NGC 4178) fall entirely within region III, exhibiting 
similar ratios to HII nuclei.

To summarize, with the exception of NGC 4178, the MIR spectra of all our 
sample galaxies strongly suggest that they do not contain AGN.

\begin{figure}[]

\noindent{\includegraphics[width=9cm]{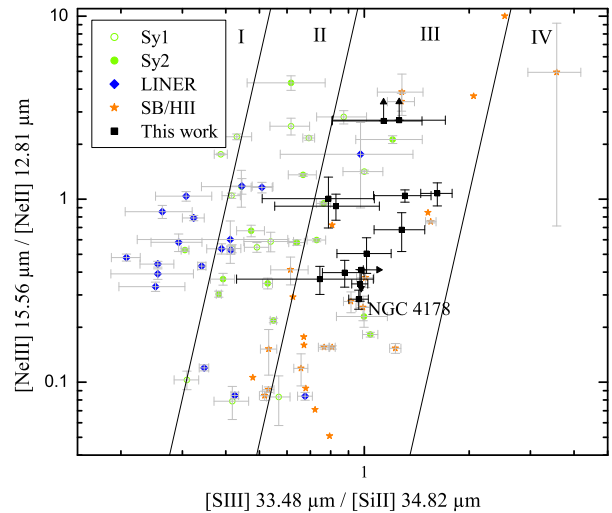}}
\caption[]{Mid-infrared diagnostic diagram used to separate AGNs from 
starbursts and "HII" galaxies from Dale et al. (2006).  Galaxies included 
are from Dale et al. (2009), Tommasin et al. 2008, Verma et al. (2003), and 
our sample of bulgeless galaxies, which are indicated by the filled squares.
  Regions I-IV are delineated by the solid line, as proposed by Dale et al. 
2006, demonstrating that no known AGNs fall in regions III and IV, where all 
of our bulgeless galaxies are located.}
\end{figure}

\subsection{ Density of the Ionized Gas}

Abundance-independent density estimates have been obtained using infrared 
fine-structure transitions from like ions in the same ionization state with 
different critical densities.  The density diagnostic available in the IRS 
spectra of our objects are the [SIII]18.71$\mu$m and  33.48 ~$\mu$m lines 
(where n$_{crit}$ $\sim 1.5 \times10^4 {\rm~cm^{-3}}$, and 4.1 $\times
10^3 {\rm~cm^{-3}}$, respectively, where n$_{crit}$ = 
A$_{ul}$/$\gamma$$_{ul}$, with A$_{ul}$ the Einstein A coefficient 
and $\gamma$$_{ul}$ the rate coefficient for collisional de-excitation 
from the upper to the lower level).  The results are largely unaffected 
by the shape of the ionizing continuum. However, in Dudik et al. (2007), 
we showed that since the [SIII] emission is generally extended and the LH 
slit is larger than the SH slit, any analysis derived using line fluxes 
obtained from apertures of different sizes is ambiguous for most nearby 
galaxies. Furthermore, differential extinction towards the emitting gas 
in very obscured sources will also affect the line flux ratio, resulting in 
further ambiguities in the interpretation of the ratio.  
Nonetheless, for comparative purposes, we list in Table 4 the 
[SIII]18.71~$\mu$m/ [SIII] 33.48 ~$\mu$m line flux ratios for our sample of 
galaxies. .  From Table 4, we see that for the galaxies in which both lines 
are detected, the line flux ratio ranges from 0.5 to 3.6, with an average 
value of 1.6. As shown in Dudik et al. (2007), at the distances for most of 
our sources, the [SIII] emission will likely extend beyond the SH aperture, 
resulting possibly in an artificially {\it lower} 
[SIII]18.71~$\mu$m/ [SIII]33.48 ~$\mu$m line flux ratio. As a comparison, 
the {\it aperture-matched} [SIII]18.71~$\mu$m/ [SIII] 33.48 ~$\mu$m line 
flux ratio in the SINGS sample of 75 galaxies from Dale et al. (2009) ranges 
from 0.3 to 2, with an average value of 0.8, significantly {\it lower} than 
the values listed in Table 4.  Indeed, 97\% of the SINGS sample has line 
flux ratios below the average value found in our sample.  Since aperture 
affects should in principle lower the line flux ratios in our sample 
compared to the aperture-matched values in Dale et al. (2009), the 
higher [SIII]18.71~$\mu$m/ [SIII] 33.48 ~$\mu$m line flux ratios found in 
our sample of bulgeless galaxies possibly implies higher gas densities 
toward the [SIII]-emitting gas. The gas densities derived using the models 
from Dudik et al. (2007) for a gas temperature of $T = 10^4~K$ (obtained 
using the collision strengths from Tayal \& Gupta (1999) and radiative 
transition probabilities from Mendoza \& Zeippen 1982) range from 
$\sim 100~{\rm cm^{-3}}$ - 3 $\times10^3{\rm~ cm^{-3}}$ for our sample.  We 
emphasize that the ambiguities inherent in the use of the [SIII] ratio, 
particularly significant in nearby galaxies, preclude us from making 
definitive statements about the actual gas densities in the ionized gas in 
our sample.  Nevertheless, the [SIII] ratios appear to imply higher ionized 
gas densities compared with standard AGN and normal/starburst galaxies of 
earlier Hubble type.  We note that Dale et al. (2009) find that the average 
[SIII] ratios is independent of whether the region probed is a star-forming 
or AGN environment.

\section {The AGN in NGC 4178}

\subsection{Line Fluxes and Spectral Line Fits }

In Figure 5, we show the spectra extracted from the full SH aperture near 
13~\micron, 14~\micron, and 15~\micron, showing three emission lines from 
different ionization states of Neon.  As can be seen, there is a clear 
detection ($\sim$ 6$\sigma$) of the [NeV]~14.32~\micron\ line, providing 
strong evidence for an AGN in this galaxy.  The [NeV] line is not resolved 
with the R=600 {\it Spitzer} SH spectral resolution.  The [NeV]/[NeII] line 
flux ratio for NGC 4178 is 0.12, within the range but at the low end of the 
values observed in standard AGNs as discussed above.  The [NeV]/[NeII] line 
flux ratio in NGC 4178 is a factor of $\sim$ 2 higher than the same ratio in 
NGC 3621, our previously discovered Sd galaxy with [NeV] emission.  The 
infrared spectra (S07, Abel \& Satyapal 2008) together with newly acquired 
additional multiwavelenth observations (Barth et al. 2009; Gliozzi et al. 
2009) have made the case for an AGN in NGC 3621 secure.  It is thus very 
likely that NGC 4178 also harbors an AGN, and that it is possibly slightly 
more energetically significant than the one in NGC 3621.  The upper limit 
to the [OIV]/[NeII] line flux ratio in NGC 4178 is 0.1, still within range 
of the observed values in standard AGNs as discussed above.  As in the case 
of NGC 3621, the low [NeV]/[NeII] and [OIV]/[NeII] line flux ratios in 
NGC 4178 suggests that the {\it Spitzer} spectrum is dominated by regions of 
star formation, and that there is significant contamination of the lower 
ionization emission lines from star formation within the {\it Spitzer} 
aperture.

The [NeV] 14.32~\micron\ luminosity observed from NGC 4178 is 
8.23$\times$10$^{37}$ ergs s$^{-1}$, slightly higher than the value 
observed in NGC 3621 (5$\times$10$^{37}$ ergs s$^{-1}$; S07) and almost 3 
orders of magnitude lower than the median value observed in standard AGNs 
(see Section 3.1.1).  It is also on the low end of the luminosities observed 
in other recently discovered late-type galaxies showing [NeV] emission 
presented in S08.  Like NGC 3621, NGC 4178 likely harbors a very weak AGN.

\begin{figure*}[htbp]
\begin{center}
\begin{tabular}{cc}
  \includegraphics[width=0.45\textwidth]{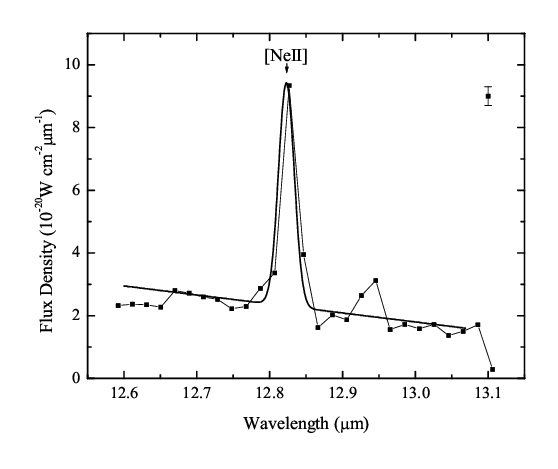} &
  \includegraphics[width=0.45\textwidth]{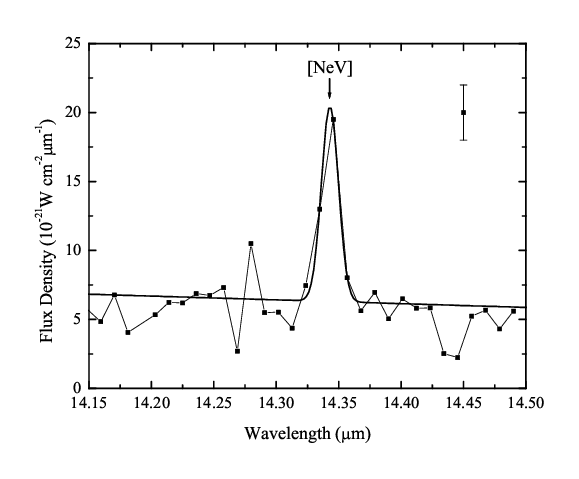} \\
  \multicolumn{2}{c}{\includegraphics[width=0.45\textwidth]{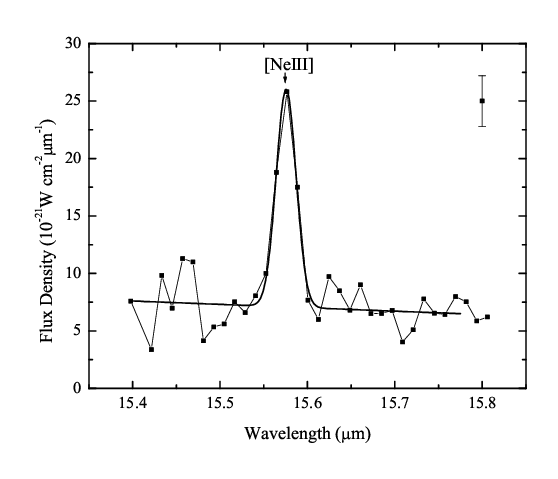}} \\
\end{tabular}
\end{center}
\caption[]{IRS SH spectra of NGC 4178 showing the detections of the [NeII]  
12.81~\micron, [NeV] 14.32~\micron, and [NeIII] 15.56~\micron fine structure 
lines. Representative error bars are displayed in the upper right corner of 
each plot.}
\end{figure*}

\subsection{Other Evidence}
There does not appear to be any previously published evidence for an AGN in 
NGC 4178.  This source was not observed by {\it Chandra} or {\it XMM-Newton}.
 The galaxy was not detected by Einstein and the upper limit to the X-ray 
luminosity,  ( L$_{\rm X}$ $\sim$ 2.5$\times$10$^{40}$ ergs 
s$^{-1}$ ; Fabbiano, Kim, \& Trinchieri 1992) is consistent with a low 
luminosity AGN (e.g. Ho et al. 2001, Dudik et al. 2005). There is no 
evidence of a central source at radio wavelengths.  The radio emission at 
2.8, 6.3 and 20 cm peaks 55 " away from the optical center, and is 
associated with optically bright knots (Niklas et al. 1995a). There is no 
evidence of an excess in the radio-FIR correlation as is seen in AGN 
(Niklas et al. 1995b) and the radio spectral index is typical of star 
forming galaxies (Vollmer et al. 2004). The [NeV] detection reported in this 
work is the first observation suggesting the presence of an AGN.

In Abel \& Satyapal (2008), we used the spectral synthesis code CLOUDY to 
model the emission line spectrum from gas ionized by both an input AGN 
radiation field and a young starburst.  In the case of NGC 3621, we showed 
that the MIR spectrum cannot be replicated unless 30-50\% of the bolometric 
luminosity within the {\it Spitzer} IRS aperture is due to an AGN.  In 
Figure 6, we show the predicted [NeV]14.3$\mu$m/[NeII]12.8$\mu$m flux ratio 
versus the [OI]/ H$\alpha$ and [SII]/ H$\alpha$ optical line flux ratios 
based on the models from Abel \& Satyapal (2008) for varying values of the 
ionization parameter (U; the dimensionless ratio of ionizing flux to gas 
density) and AGN luminosity contribution. We display only a narrow range of 
ionization parameters that generate line flux ratios within the range 
observed in NGC 4178 and NGC 3621.  A more extensive grid of theoretical 
calculations, with all standard optical line flux ratios plotted, is 
presented in Abel \& Satyapal (2007).  As can be seen from Figure 6, the 
MIR and optical emission line spectra of NGC 4178 cannot be replicated 
with a pure starburst ionizing radiation field.  An AGN contribution of 
$\sim$ 30-90\% is required. 

\begin{figure}[htbp]
\begin{center}
  \includegraphics[width=0.45\textwidth]{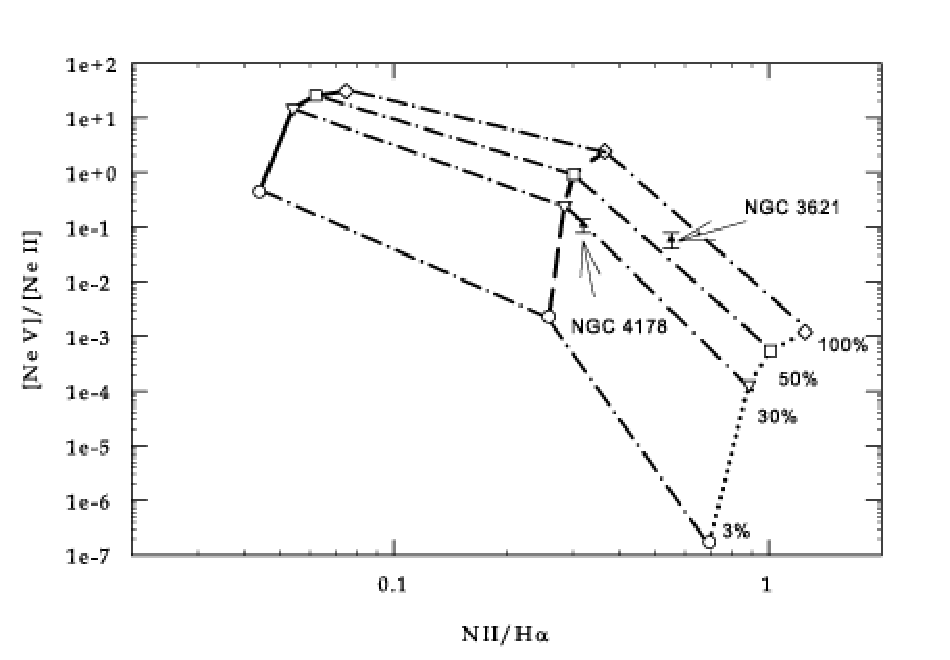} 
 \includegraphics[width=0.45\textwidth]{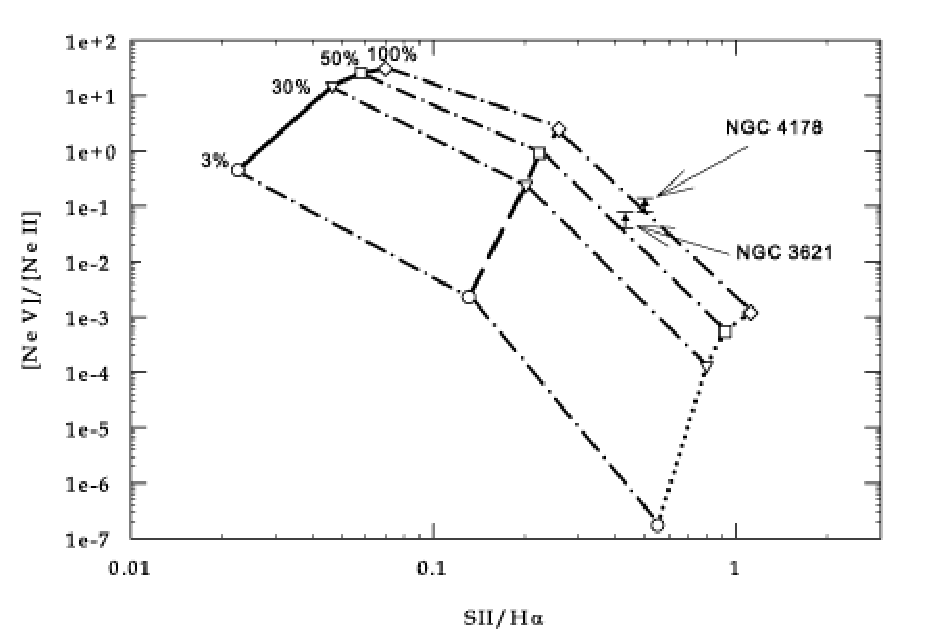} 
\end{center}
\caption[]{The [NeV]14.3$\mu$m/[NeII]12.8$\mu$m line flux ratio versus the 
optical (a) [NII]/ H$\alpha$ flux ratio and (b) the [SII]/ H$\alpha$ flux 
ratio from the models from Abel \& Satyapal (2008).  The solid, dashed, and 
dotted line display model results for ionization parameters of - 1.5, -2.5, 
and -3.5, respectively.  The dashed-dotted lines show the fraction of the 
total luminosity due to the AGN.  The line attached to the circles represent 
3\% AGN, inverted triangles 10\% AGN, squares 30\% AGN, diamond 50\% AGN, 
and triangles 100\% AGN, as indicated in the Figure.  The line flux ratios 
for NGC 4178 and NGC 3621 are shown.  We note that the optical line flux 
ratios for NGC 4178 are taken directly from Table 4 in H97.  The 
[NII]/H$\alpha$ flux ratio for NGC 3621 is taken from the large aperture 
(20"$\times$20") measurements from Dale et al. 2006.  The 
[SII]/H$\alpha$ measurement for NGC 3621 is from the higher spatial 
resolution (1~\arcsec) observations from Barth et al. (2009).}
\end{figure}

Although it is clear that gas photoionized solely by even the youngest 
starburst ionizing radiation field cannot simultaneously reproduce the 
optical and mid-infrared spectrum of NGC 4178, the possibility that the 
[NeV] emission originates from shocked gas associated with a 
starburst-driven superwind (e.g Veilleux, Cecil, \& Bland-Hawthorn 2005) 
needs to be explored.  Since the large aperture of the Spitzer IRS modules 
precludes us from using morphological arguments to rule out shocks, we 
consider whether the combined optical and mid-infrared fine structure line 
ratios are consistent with radiative shock models.  Using the extensive 
grid of models from the most recent MAPPINGS III shock and photoionization 
code (Allen et al. 2008), we find that the optical spectrum of NGC 4178 
cannot be reproduced for most of the parameter space they studied.  In fact, 
the only shock models which can simultaneously reproduce the observed 
[O III]/H$\beta$, [N II]/H$\alpha$, and [S II]/H$\alpha$ line ratios for 
NGC 4178 are for high-density (n(H) = 1000 cm-3) shocks (Allen et al. 2008; 
Figure 22a).  However, at these densities, the mid-infrared spectrum, in 
particular the [Ne III]/[Ne II]  is 1-2 orders of magnitude higher than the 
observed value of 0.28 for NGC 4178.  Thus, the mid-infrared spectrum and 
the optically "normal" spectrum of NGC 4178 cannot be simultaneously 
replicated by any shock model. Although high density radiative shocks may 
play a role in the emission line spectrum of NGC 4178 (and other AGN), it 
appears that the combined optical and mid-infrared spectrum cannot be 
produced without an AGN radiation field.  Follow-up {\it Chandra} 
observations are crucial to confirm the presence of the AGN and constrain 
its location.

\subsection { Bolometric Luminosity and Black Hole mass limit}

We can obtain an order of magnitude estimate of the bolometric luminosity 
of the AGN in NGC 4178 using the [NeV] line luminosity. Assuming that the 
line emission arises exclusively from the AGN, we follow the procedure 
adopted by S07 and S08 to estimate the nuclear bolometric luminosity of 
the AGN.   Using the tight correlation between the [NeV] 14~$\mu$m line 
luminosity and the AGN bolometric luminosity found in a large sample of 
standard AGN (Equation 1 in S07), the AGN bolometric luminosity of NGC 4178 
is $\sim$ 8$\times$10$^{41}$ ergs s$^{-1}$, slightly greater than the 
estimate for the AGN bolometric luminosity of NGC 3621 (S07).  This 
estimate assumes that the relationship between the [NeV] 14~$\mu$m line 
luminosity and the bolometric luminosity established in more luminous AGN 
(see S07) extends to the lower [NeV] luminosity range characteristic of 
NGC 4178 and other late-type galaxies.  The nuclear bolometric luminosities 
of the AGNs discovered in the late-type galaxies from S08 range 
from $\sim 3\times10^{41}{\rm~ ergs s^{-1}}$ to $\sim 2\times10^{43}{\rm~
 ergs s^{-1}}$, with a median value of 
$\sim 9\times10^{41}{\rm~ ergs s^{-1}}$.  As can be seen, the luminosity of 
the AGN in NGC 4178 is typical of other recently discovered AGN in low-bulge 
galaxies.

If we assume that the AGN is radiating below the Eddington limit, we can 
estimate the lower limit to the mass of the black hole based on the AGN 
bolometric luminosity estimate.  The Eddington mass estimate in NGC 4178 is 
$\sim 6\times10^3{\rm~M_{\odot}}$ , well within the range of lower mass 
limits found in other late-type galaxies with AGN (S07, S08).  There appears 
to be a nuclear star cluster in NGC 4178 (see section 7.1).  However, there 
are no measurements of the central velocity dispersion.  We therefore cannot 
determine if the lower mass limits derived for the black hole mass are 
incompatible with the M$_{BH}$-$\sigma$ relation, assuming a linear 
extrapolation to the low mass range.

\subsection {Comparison to other AGNs in Bulgeless Galaxies}

NGC 4178 is one of less than a handful of completely bulgeless disk galaxies 
showing evidence for an AGN.  The best-studied definitively bulgeless disk 
galaxy with an AGN is the galaxy NGC 4395, which shows the hallmark 
signatures of a type 1 AGN (e.g. Filippenko \& Ho 2003; Lira et al. 1999; 
Moran et al. 1999).  The bolometric luminosity of the AGN is 
$\sim 10^{40}{\rm~ergs s^{-1}}$ (Filippenko \& Ho 2003), almost two orders 
of magnitude lower than the estimated bolometric luminosity of the AGN in 
NGC 4178.  The estimated bolometric luminosity of the AGN in NGC 3621 is a 
factor of 1.6 less (S07) than that of the AGN in NGC 4178, making NGC 4178 
the most luminous AGN in an Sd galaxy currently known.  The black hole mass 
of NGC 4395, determined by reverberation mapping, is 
M$_{BH}=(3.6\pm 1.1)\times10^5 {\rm~M_{\odot}}$ (Peterson et al. 2005).  
This value for the black hole mass would be consistent with the lower limit 
of the black hole mass derived for NGC 4178 of 6$\times$10$^3$M$_{\odot}$ if 
the AGN is accreting at a high rate.  Our recent X-ray observations, when 
combined with our {\it Spitzer} observations, suggest that the black hole 
mass in NGC 3621 is $\sim$ 2$\times$10$^4$M$_{\odot}$ and that it is 
accreting at a high rate (L$_{bol}$/L$_{Edd}$ $>$ 0.2) (Gliozzi et al. 2009).
  If NGC 4178 is similar to NGC 3621, then its black hole mass is comparable 
to that in NGC 4395 and is inline with those inferred for nuclear black 
holes in other pseudobulge galaxies (Greene \& Ho 2007).

\section {Mid-Infrared AGN Detection Rate in Sd Galaxies}

In S08, we showed that optical studies significantly miss AGN in late-type 
galaxies.  From the H97 sample, out of the full sample of 486 galaxies, 207 
are of Hubble type Sbc or later, and only 16 (8\%) are optically classified 
as AGN.  Using MIR diagnostics, we demonstrated that the AGN detection rate 
in optically normal disk galaxies of Hubble Type Sbc or later, is 
$\sim$ 30\%, implying that the overall fraction of late-type (Sbc or later) 
galaxies hosting AGNs is possibly more than 4 times larger than what optical 
spectroscopic studies indicate.  Although it is now clear that AGNs do 
reside in a significant number of late-type galaxies, virtually all of the 
newly discovered AGNs are in galaxies with Hubble type of Scd or earlier. 
Prior to the current work, there were only a handful of Sd galaxies observed 
by the high resolution modules of {\it Spitzer's} IRS, precluding us from 
determining based on MIR diagnostics the true AGN fraction in galaxies with 
essentially {\it no} bulge.

In Figure 7, combining our current sample with that from S08, we show the 
AGN detection fraction in optically normal galaxies as a function of Hubble
 type. Since the sensitivity of the observations varied across the sample, 
we also indicate with a downward arrow in Figure 6 the number of galaxies 
with [NeV] 14~$\mu$m 3$\sigma$ line sensitivity of 10$^{38}$ ergs s$^{-1}$  
or better.  As can be seen, all of the Sd/Sdm galaxies were observed with 
the highest sensitivity.  There are a total of 22 Sd/Sdm galaxies observed 
by {\it Spitzer} IRS and only 1 with a [NeV] detection (NGC 4178).  
Figure 7 shows that the AGN detection rate in optically normal galaxies 
drops dramatically for pure disk galaxies, with a detection rate of only 
4.5\%.  From the H97 sample, out of the full sample of 486 galaxies, 
excluding interacting and irregular galaxies, 26 are of Hubble type 
Sd/Sdm and only one is optically identified as an AGN (NGC 4395).  With 
the discovery of only one additional AGN in an Sd galaxy from the H97 
sample based on MIR diagnostics, the overall detection rate of AGN in 
pure disk galaxies is only $\sim$ 8\%, significantly lower than the 
detection rate in late-type galaxies with some bulge component.  Our 
study thus shows that AGNs in pure disk galaxies do not appear to be 
hidden but are indeed truly rare.

\begin{figure}[]
\noindent{\includegraphics[width=9cm]{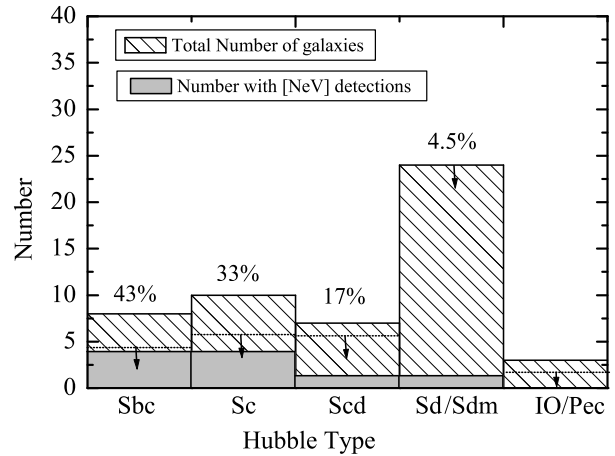}}
\caption[]{The distribution of Hubble types for the current sample 
combined with that from S08.  The galaxies with [NeV] detections are 
indicated by the filled histogram.  Since the sensitivity of the 
observations varied across the sample, we also indicate with a downward 
arrow in Figure 6 the number of galaxies with [NeV] 14~$\mu$m line 
sensitivity of 10$^{38}$ ergs s$^{-1}$or better. As can be seen the AGN 
detection rate in optically normal galaxies drops dramatically for 
galaxies with no bulge component.}
\end{figure}

\section{Demographics of Late-type Galaxies with AGNs}

\subsection{A Nuclear Star Cluster in NGC 4178?}

With the discovery of an AGN in NGC 4178, there are now only 3 known 
AGN in Sd galaxies. As mentioned earlier, the two other Sd galaxies with
AGNs, NGC 4395 and NGC 3621, both have prominent NSCs. In Figure 8, we
show the {\it HST} NICMOS image of NGC 4178 (B\"oker et al. 1999). The 
image reveals a prominent source close to the apparent photocenter, 
indicated by an
arrow. As will be discussed in Section 8, the luminosity of this source 
is consistent with that of a NSC. Unfortunately, the NICMOS image is not 
centered well, and thus does not allow an unambiguous determination of the 
photocenter.

However, visual inspection and a cursory isophote analysis over the limited 
NICMOS field of view suggest that the location of the NSC is indeed
 consistent with that of the photocenter of NGC 4178. Because the spatial
resolution of the {\it Spitzer} data precludes us from determining the
spatial location of the [NeV] peak to determine whether it coincides
with the NSC, follow-up {\it Chandra} observations are crucial to 
confirm that the AGN indeed resides within the putative NSC. If we 
assume for the moment that the prominent central source in the NICMOS 
image is a NSC, then one can infer that all known AGNs in Sd galaxies 
reside in NSCs, possibly suggesting that in the absence of
any bulge, an NSC is required for an AGN to be present.

We can estimate a rough mass for the NSC in NGC 4178 if we assume the 
average I-band M/L ratio for NSCs in late-type galaxies with measured 
dynamical masses (Walcher et al. 2005). Based on the rough I-band magnitude 
estimate (see Section 8), the nuclear cluster mass in NGC 4178 is 
$\sim 0.5\times10^6{\rm~M_{\odot}}$, comparable to the nuclear cluster mass 
in NGC 4395 (Seth et al. 2008) and an order of magnitude less than the one 
in NGC 3621 (Barth et al. 2009).  Seth et al. (2008) find that in cases 
with known nuclear cluster and black hole masses, the ratio of the black 
hole mass to nuclear cluster mass, M$_{BH}$/M$_{NC}$, ranges from 0.1-1.  
This ratio is $\sim$ 1/3 in NGC 4395. Using the lower limit to the black
 hole mass for NGC 4178, we find M$_{BH}$/M$_{NC} > 10^{-2}$, which 
is possibly consistent with the ratio found in galaxies with measured 
nuclear cluster and black hole masses (Seth et al. 2008).

\begin{figure}[]
\noindent{\includegraphics[width=9cm]{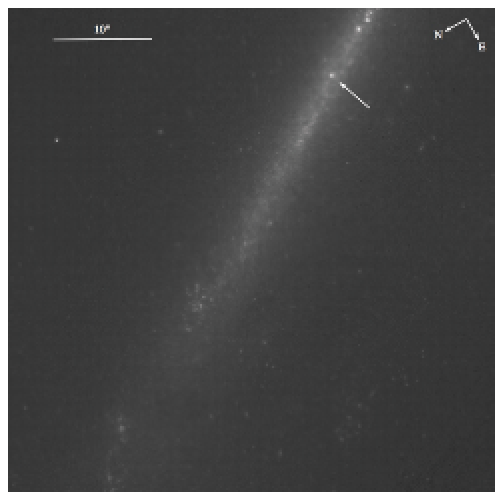}}
\caption[]{{\it HST} NICMOS H-band image of NGC 4178. The location of 
the NSC is indicated by the arrow and is consistent with the apparent 
photocenter. The field of view of the image is 51"$\times$51".}
\end{figure}

\subsection{AGNs in Late-type Bulges}

It is well established that AGN are common in early-type galaxies and that 
there is a trend of increasing AGN activity with bulge mass (e.g. H97; 
Kauffman et al. 2003).  Based on our study and other recent studies, it is 
also now clear that AGN do exist in late-type galaxies that lack a classical 
bulge and that they are significantly more common than previously thought 
(S07;S08; Greene, Ho, \& Barth (2009); Shields et al. 2008; Ghosh et al. 
2008; Barth et al. 2008; Dewangan et al. 2008; Desroches \& Ho 2009). 
Late-type galaxies are often characterized by so-called pseudobulges, with 
exponential surface brightness profiles similar to disks rather than 
classical bulges.  These pseudobulges are thought to have formed from 
quiescent secular processes within the host galaxy, in contrast to the 
violent merger-driven events thought to have formed classical bulges (see 
review in Kormendy \& Kennicutt 2004).  It is therefore relevant to ask 
whether the incidence and properties of BHs in late-type galaxies are 
related to the presence and properties of pseudobulges.

Combining the samples from S08, and including NGC 3621 (S07), and this work, 
there are a total of 52 nearby optically normal late-type galaxies in which 
the presence or absence of AGN has been determined by MIR diagnostics.  With 
a total of 9 AGN in this combined sample, we can attempt to investigate the 
relationship between AGN activity and the host galaxy properties for 
galaxies of Hubble type Sbc or later.  We emphasize, that the AGNs in this 
sample were previously unknown based on optical spectroscopic studies.  The 
study of the demographics of these newly discovered AGNs allows us to 
investigate whether there are any trends in AGN activity with host galaxy 
properties in late-type galaxies that were previously unseen in studies 
based on optical observations.

For those 13 galaxies in our current sample with existing {\it HST} imagery, 
we have used the archival {\it HST} data to establish the presence or 
absence of a nuclear star cluster (see Table 1).  None of these {\it HST} 
images show any evidence for a bulge component, confirming our selection 
criterium, and we therefore assume the remaining 5 galaxies in our sample 
are bulgeless as well.
For the remaining galaxies in our combined {\it Spitzer} sample, we 
searched the literature for all information on the structural properties of 
the host galaxies. Of the 52 late-type galaxies, 32 had published surface 
brightness profile fits to characterize the bulge properties 
(Knapen et al. 2003; Laurikainen et al. 2004; Scarlata et al. 2004; 
Dong \& De Robertis 2006; Drory \& Fisher 2007). We point out that 
several of the published bulge properties are based on ground-based 
imagery with low spatial resolution, which can significantly compromise 
the inferred bulge parameters.  In addition, the presence or absence of 
NSCs has not been investigated in all galaxies.  We therefore do not carry 
out a quantitative analysis of the relationship between bulge parameters 
and AGN presence and properties. Instead, we investigate whether there are 
any evident {\it trends} between the host galaxy properties and the 
incidence of AGN activity.  Most of the galaxies in our previous sample 
(S08) have pseudobulges or weak classical bulges and four have identified 
NSCs.  We point out that the NSC presence has not been investigated in the 
majority of galaxies in the sample and it is likely that most of these 
late-type galaxies do have NSCs.  Of the 8 AGNs in our S08 sample, 6 have 
published surface brightness profile fits. Amongst these sources, 2 are 
reported to have classical bulges (NGC 3367, NGC 4414; Dong \& De Robertis 
2006, Laurikainen et al. 2004, respectively), and the remaining 4 are 
reported to have surface brightness profiles consistent with pseudobulges 
(NGC 3938, NGC 4321, NGC 4536, and NGC 5055; Dong \& De Robertis 2006, 
Scarlata et al. 2004, Drory \& Fisher 2007).  One of the AGN galaxies is 
reported in the literature to host a NSC (NGC 4321; Knapen et al. 1995), 
but the presence of absence of a NSC in the remaining AGN galaxies has not 
been established. Although the small sample size and the ambiguity in the 
published bulge properties precludes us from conducting a quantitative 
investigation of the relationship between bulge parameters and nuclear 
cluster presence and AGN activity, it appears that most AGNs in late-type 
galaxies reside in galaxies with pseudobulges or weak classical bulges.  
It is also clear that NSCs and AGNs coexist, consistent with the findings 
from Seth et al. (2008) in more massive galaxies spanning a wide range of 
Hubble types.

Most importantly, our current study robustly shows that AGNs are extremely 
rare in bulgeless galaxies and that for the few cases where one does exist 
(NGC 3621, NGC 4178, and NGC 4395), there is a prominent NSC. These 
findings possibly suggest that if there is no bulge of any kind in a 
galaxy, the galaxy {\it must have a NSC} in order to host an AGN.

\section {Is NGC 4178 Special?}

Our findings demonstrate that AGN are truly rare in bulgeless galaxies.  
An important question to then ask is what distinguishes bulgeless disk 
galaxies {\it with} AGN from those {\it without} AGN?  Is the presence 
and properties of the black hole in any way related to the properties 
of the host galaxies in cases where there truly is {\it no} bulge?  
With 18 bulgeless disk galaxies in our sample and only one AGN 
(NGC 4178), the question arises whether NGC 4178 is in some way special 
in our sample. Of course the absence of an AGN does not imply the absence 
of a quiescent massive black hole in any of the galaxies in our sample. 
The only dynamical study that rules out the presence of a massive black 
hole (M$_{BH} < 1500 {\rm~M_{\odot}}$) in a bulgeless disk galaxy was 
carried out for the nearby galaxy M33 (Gebhardt et al. 2001).

We investigated whether NGC 4178 is unique in any way in its basic host 
galaxy properties listed in Table 1.  The total estimated galaxy mass for 
NGC 4178 is $\sim 10^8{\rm~M_{\odot}}$, on the high end but not the highest 
of the galaxies in our sample (see Figure 1).  From Table 1, we can see that 
the estimated HI mass is high but again comparable to or lower than several 
other galaxies in the sample, implying that the disk mass does not appear to 
be related to the presence or absence of an AGN in bulgeless disk galaxies.  
Similarly, the inclination-corrected HI rotational amplitude of NGC 4178 is 
high but not the highest in the sample (see Table 1), suggesting that the 
total dark matter mass is not the determining factor in whether or not a 
bulgeless disk galaxy hosts an AGN.  Finally the nuclear SFR in NGC 4178 
listed in Table 1, estimated using the exintction-corrected H$\alpha$ 
luminosity, is only slightly larger than the median value for the entire 
sample, but well below the highest value in the sample.  There is thus no 
clear indication that the basic host galaxy properties in NGC 4178 are 
exceptional in any way compared to the rest of our sample.

If the basic host galaxy properties in NGC 4178 do not distinguish 
themselves from the rest of sample, we can ask whether the nuclear cluster 
properties do.  Using the {\it HST} NICMOS image of NGC 4178, we estimate a 
magnitude of m$_{H}$=18.24 using a circular aperture of 3 pixels centered 
on the putative NSC. We point out that in this analysis, we are assuming 
that the AGN in NGC 4178 is coincident with the putative NSC - an assumption 
which is not possible to confirm with the poor spatial resolution of the 
{\it Spitzer} data. Follow-up high spatial resolution {\it Chandra} 
observations are crucial to confirm this hypothesis. Using the latest 
stellar population synthesis models from Bruzual \& Charlot (2009), the 
I-H color for single age population older than $10^8$ years, assuming 
solar metalicity and a Chabrier (2003) initial mass function does not 
exceed 1.7.  This means that the nuclear cluster in NGC 4718 has an absolute 
I-band magnitude of M$_{I}$ $\sim$ -11, typical of the I-band magnitudes of 
nuclear clusters in the extensive sample of nuclear clusters in late-type 
galaxies from B\"oker et al. (2002) (see their Figure 5).  We note that we 
have not made any extinction correction, which could be significant in an 
edge-on galaxy such as NGC 4178.  We also assumed an old stellar population.
  Since NGC 4178 shows prominent low ionization emission lines indicating 
the presence of young stars, the nuclear cluster color might be bluer than 
assumed above, resulting in an underestimate of the I-band luminosity.  
Finally, since the AGN is weak and hidden at optical wavelengths, we have 
assumed that the AGN contribution to the central luminosity in the H-band 
is negligible.  Our estimate of the nuclear cluster luminosity should 
therefore be considered approximate. However, it does appear based on 
this rough estimate, that the nuclear cluster luminosity in NGC 4178 is 
typical of a nuclear cluster in late-type galaxies. There is thus no clear 
indication that NGC 4178 distinguishes itself from the rest of our sample 
of disk galaxies both in terms of the overall galaxy properties or its 
nuclear cluster luminosity.  The recipe for forming and growing a massive 
black hole in a truly bulgeless disk galaxy is still unknown.

\section{Summary and Conclusion}

We conducted a MIR spectroscopic investigation of 18 completely bulgeless 
disk galaxies showing no signatures of AGN in their optical spectra in 
order to search for low luminosity and/or embedded AGN.  This is the first 
systematic search for weak or hidden AGN in a statistically significant 
sample of esssentially bulgeless disk galaxies. The primary goal of our 
study was to determine the incidence of AGNs in galaxies in the absence 
of a significant bulge.  Our high resolution {\it Spitzer} spectroscopic 
observations reveal that while AGNs in galaxies with pseudobulges or weak 
classical bulges are significantly more common than once thought, AGNs in 
truly bulgeless disk galaxies are exceedingly rare.  Our main results are 
summarized below:

\begin{enumerate}

\item We detected the high ionization [NeV] 14.3~$\mu$m emission line in 
only one out of the 18 galaxies in the sample, providing strong evidence 
for an AGN in this one source. This galaxy, NGC 4178, is a nearby 
(d=16.8 Mpc) edge-on disk galaxy with optical emission line ratios in the
 normal star 
formation regime, indicating that there is absolutely no hint of an AGN 
based on its optical spectrum.

\item With the exception of NGC 4178, none of the galaxies in the sample 
shows any evidence in the MIR for a weak or embedded AGN, suggesting that 
they lack AGN.  Instead, most galaxies show signs of active star formation 
and possibly higher ionized gas densities than galaxies of earlier Hubble 
type.

\item Our work suggests that the AGN detection rate based on MIR diagnostics 
in late-type optically normal galaxies is high (30-40\%) in galaxies of 
Hubble type Sbc and Sc but drops drastically in Sd/Sdm galaxies (4.5\%). 
Our observations confirm that AGNs in completely bulgeless disk galaxies 
are not hidden in the optical but truly are rare.

 \item The AGN bolometric luminosity of NGC 4178 inferred using our [NeV] 
line luminosity is $\sim$ 8$\times$10$^{41}$ ergs s$^{-1}$, a factor of 1.6 
times greater than the estimated bolometric luminosity in the Sd galaxy 
NGC 3621, and almost two orders of magnitude greater than the AGN bolometric 
luminosity of NGC 4395, the best-known AGN in an Sd galaxy.  This makes the 
AGN in NGC 4178 the most luminous known in a bulgeless disk galaxy.  
Assuming that the AGN is radiating below the Eddington limit, this c
orresponds to a lower mass limit for the black hole of 
$\sim 6\times 10^3{\rm~M_{\odot}}$. There are no published measurements of 
the central stellar velocity dispersion.  It is therefore unknown if the 
lower mass limit for the black hole in NGC 4178 violates the 
M$_{\rm BH}$-$\sigma$ relation established in early-type galaxies.

\item NGC 4178 is now one of only 3 known Sd galaxies showing evidence for 
an AGN.  {\it HST} images of this galaxy suggests that it has a prominent 
NSC, similar to the ones seen in the other two known Sd galaxies with AGNs 
(NGC 3621, NGC 4395).  If follow-up {\it Chandra} observations confirm that 
the AGN is coincident with the putative NSC in this source, this finding 
suggests that if there is {\it no} bulge of any kind in a galaxy, the 
galaxy must have a NSC in order to host an AGN.

\item We find that NGC 4178 is not exceptional in our sample of 18 
bulgeless galaxies based both on its basic host galaxy properties 
(galaxy mass, disk mass, dark matter halo mass, nuclear SFR) and nuclear 
cluster properties.  The recipe for forming and growing a black hole in a 
truly bulgeless disk galaxy still remains a mystery.

\end{enumerate}

\acknowledgements
It is a pleasure to thank Rachel Dudik for her invaluable help in 
consulting with us on data analysis issues, for troubleshooting various 
software installation roadblocks, for her technical assistance in planning 
the observations, and for stimulating science discussions.  This work would 
not have been possible without her expertise in IRS data analysis.  We are 
also very grateful to the {\it Spitzer} helpdesk for numerous emails in 
support of our data analysis questions. Brian O'Halloran and Dan Watson 
were very helpful in consulting with us on data analysis procedures. This 
work is based on observations taken with the {\it Spitzer} Space Telescope, 
which is operated by JPL/Caltech under a contract with NASA. The thoughtful 
suggestions of the anonymous referee helped improve this paper. We are also 
very grateful for a fruitful discussion with Dave Alexander and Andy 
Goulding, which led us to outline more explicitly our data analysis 
procedure in Section 3.  This research has made use of the NASA/IPAC 
Extragalactic Database (NED), which is operated by the Jet Propulsion 
Laboratory, California Institute of Technology, under contract with the 
National Aeronautics and Space Administration.  SS gratefully acknowledges 
financial support from NASA grant RSA 1345391.

{}

\end{document}